\documentclass[11pt,letterpaper]{article}
\usepackage[utf8]{inputenc}
\usepackage[english]{babel}
\usepackage{amsmath}
\usepackage{amsfonts}
\usepackage{amssymb}
\usepackage{graphicx}
\usepackage[left=3cm,right=3cm,top=2cm,bottom=3cm]{geometry}
\author{Jorge Pinochet}
\title{\textbf{Brown dwarfs and the minimum mass of stars}}
\begin{document}

\author{Jorge Pinochet\\ \\
 \small{\textit{Facultad de Educación}}\\
 \small{\textit{Universidad Alberto Hurtado, Erasmo Escala 1835, Santiago, Chile.} japinochet@gmail.com}\\}

\date{}
\maketitle

\begin{center}\rule{0.9\textwidth}{0.1mm} \end{center}
\begin{abstract}
\noindent Stars form from large clouds of gas and dust that contract under their own gravity. The birth of a star occurred when a fusion reaction of hydrogen into helium has ignited in its core. The key variable that determines the formation of a star is mass. If the mass of the contracting cloud is below certain minimum value, instead of a star, a substelar object -known as a \textit{brown dwarf}- will form. How much mass is required for a star to form? This article aims to answer this question by means of a simple heuristic argument. The found value is $\sim 0.016$ solar masses, which is of the same order of magnitude as the accepted value $\sim0.08$ solar masses. This article may be useful as pedagogical material in an introductory undergraduate astronomy course.\\ \\


\begin{center}\rule{0.9\textwidth}{0.1mm} \end{center}
\end{abstract}

\maketitle

\section{Introduction}

Since ancient times, humankind has stared in amazement at the night sky and wandered about the origin of stars. Thanks to the advancement of astronomy in the last century, now we know that stars emerge from vast clouds of gas and dust that contract due to their own gravity. A star born when on its center a fusion reaction of hydrogen into helium has started, releasing huge amounts of energy. The key variable that determines if a star will form is mass. While mass does not exceed a certain value, fusion will not start and instead of a star, a substelar object will form. This object is known as a \textit{brown dwarf}\footnote{The term brown dwarf was proposed in 1975 by American astronomer Jill Tarter (1944), who is best known for her extensive work in the search of extraterrestrial intelligent life.}. How much mass is required for a star to form? In other words, how large does the initial cloud need to be to start the fusion of hydrogen into helium? Typically, an answer to this requires complex calculations, which reduces the audience to whom the topic may be accessible.\\

The main goal of this article is to develop an original argument that allows us to obtain an approximate value for the minimum mass $M_{min}$ for a star to form. Throughout the article, only elemental algebra is used to calculate $M_{min}$ along with some general notions of Classical Physics and Quantum Theory. The found value is $M_{min} \approx 0.016M_{\odot}$, where $M_{\odot}$ is the mass of the sun. This number is the same order of magnitude to the accepted value of $M_{min} \approx 0.08M_{\odot}$ (this figure is also usually expressed as a function of the Jupiter mass, $M_{J} \cong 0.001M_{\odot}$, so that $M_{\odot} \approx 80M_{j}$)\footnote{Recent research suggests that $M_{min}$ may be slightly higher [1]. However, this is an open problem that is the subject of intense debate.} [2-5]. It is worth noting that the path to achieve this result is as important as the result itself, because it allows to learn and appreciate many interesting aspects of astrophysics and stellar formation.\\

The structure of the article is as follows: In section 2, we present a wide view of the subject and a brief roadmap. In sections 3 and 4, the concept of \textit{degeneracy pressure} is introduced and from there the mass of a brown dwarf in equilibrium is calculated. In section 5 we analyze the conversion of gravitational potential energy into thermal energy and calculate the value for $M_{min}$. To close, we analyze the discovery of brown dwarfs and its relation with $M_{min}$.

\section{Stars, protostars and brown dwarfs: An overview}

A star’s formation process begins when a large cloud of dust and gas (mainly hydrogen) contracts under its own gravity; this cloud is called a \textit{molecular cloud}. Stellar formation models show that the contracting cloud will end up forming one or many spheres which physical behavior will be practically independent from its surrounding environment. These spherical structures are stellar embryos known as \textit{protostars} [3, 5].\\

When a protostar contracts, gravitational potential energy becomes translational kinetic energy of its particles\footnote{Because of the conservation of angular momentum, the contraction also gives rise to rotational kinetical energy. However, for the sake of simplicity, in this and the next sections we have ignored the rotation of the protostars.}. As a result, particles collide with each other, increasing the thermal kinetic energy [6]. Part of the thermal energy becomes electromagnetic radiation [2] (see Fig. 1). As the protostar continues to contract, increasing its density, temperature rises reaching values of $10^{3} K$. At this point, the hydrogen atoms begin to gradually ionize, until the protostar becomes a sphere composed mainly of plasma, which can be modeled as an ideal completely ionized hydrogen gas [7]: a gas composed only of free protons and electrons (having only kinetic energy).\\

\begin{figure}
  \centering
    \includegraphics[width=0.7\textwidth]{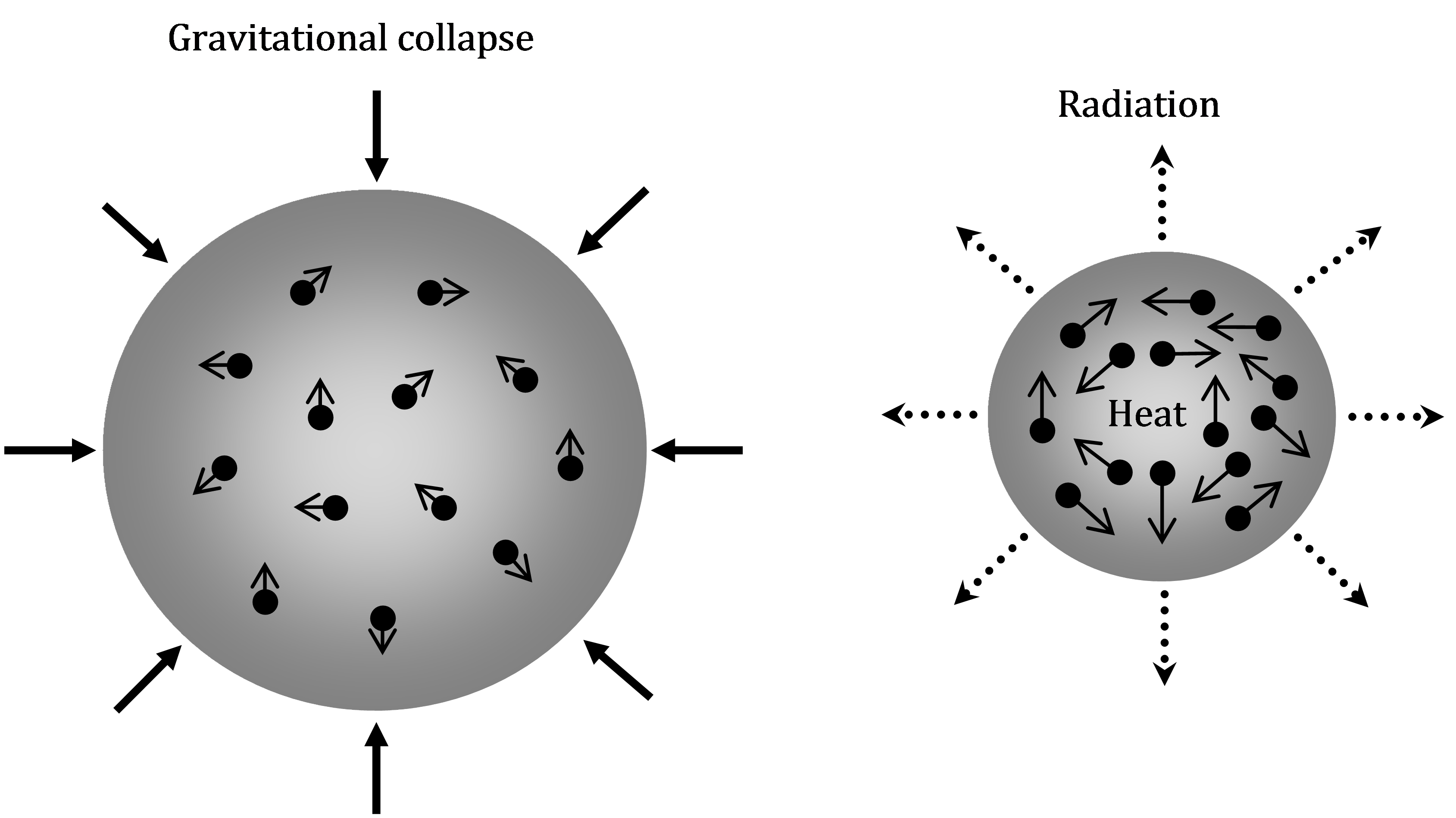}
  \caption{\small As the protostar contracts, gravitational potential energy becomes kinetic energy. This kinetic energy manifests as an increase in the average speed of the particles (represented by arrows). Part of this thermal energy becomes electromagnetic radiation.}
\end{figure}

Stellar formation models agree on a value of $M_{min} \approx 0.08M_{\odot}$ [2-5]. Therefore, if the mass reaches or exceeds this value, the process of contraction continues until the protostars center reaches the temperature required to ignite thermonuclear reactions, where hydrogen is fused into helium. This temperature is $\sim 10^{7} K$ and marks the birth of a star\footnote{Scientists have developed methods to produce fusion of hydrogen into helium in their laboratories. As in the stars, this method requires raising the temperature of a plasma to $\sim 10^{7} K$. There exists no material however that can contain plasmas at such high temperatures. To solve this problem, the most promising procedure is \textit{magnetic confinement}, wherein the hot plasma are contained in a magnetic “cage” made by strong magnetic fields which prevent the particles from escaping.} [2, 6]. These high temperatures generate an expansive pressure that can stop the gravitational collapse, allowing the star to reach a state of equilibrium, where it will remain for approximately 90\% of its life in what is called the \textit{main sequence} [8].\\

If the mass of the protostars is less than $\sim 0.08M_{\odot}$, its temperature does not reach the threshold to fuse hydrogen into helium and instead of a star, a brown dwarf is formed\footnote{Although brown dwarfs do not have a mass, temperature and pressure high enough to produce fusion of hydrogen nuclei into helium, their mass is high enough to produce deuterium fusion, also called \textit{deuterium burning}, in which a deuterium nucleus and a proton combine to form a helium-3 nucleus. The research suggests that deuterium burning turns on near a mass of 13 Jupiter masses. As additional information, this mass is often used as a criterion to distinguish brown dwarfs from planets.} [3, 4]. Due to the relatively low temperatures of these substelar objects, thermal pressure cannot stop gravitational contraction. As a result, brown dwarfs start to collapse under their own weight, gradually increasing their density. When the density is high enough, a pressure originated at the quantum level opposes further contraction and collapse. We are referring to the \textit{degeneracy pressure}, which is a product of the repulsion between identical fermions due to \textit{Pauli exclusion principle} [9] (Fig. 2). It can be shown that this pressure is associated mainly with the electrons, because the contribution of the nuclei is very low and can be ignored. The degeneracy pressure can stop the gravitational contraction completely. When this happens, the brown dwarf reaches a final state of equilibrium where it slowly cools down.\\

\begin{figure}
  \centering
    \includegraphics[width=0.4\textwidth]{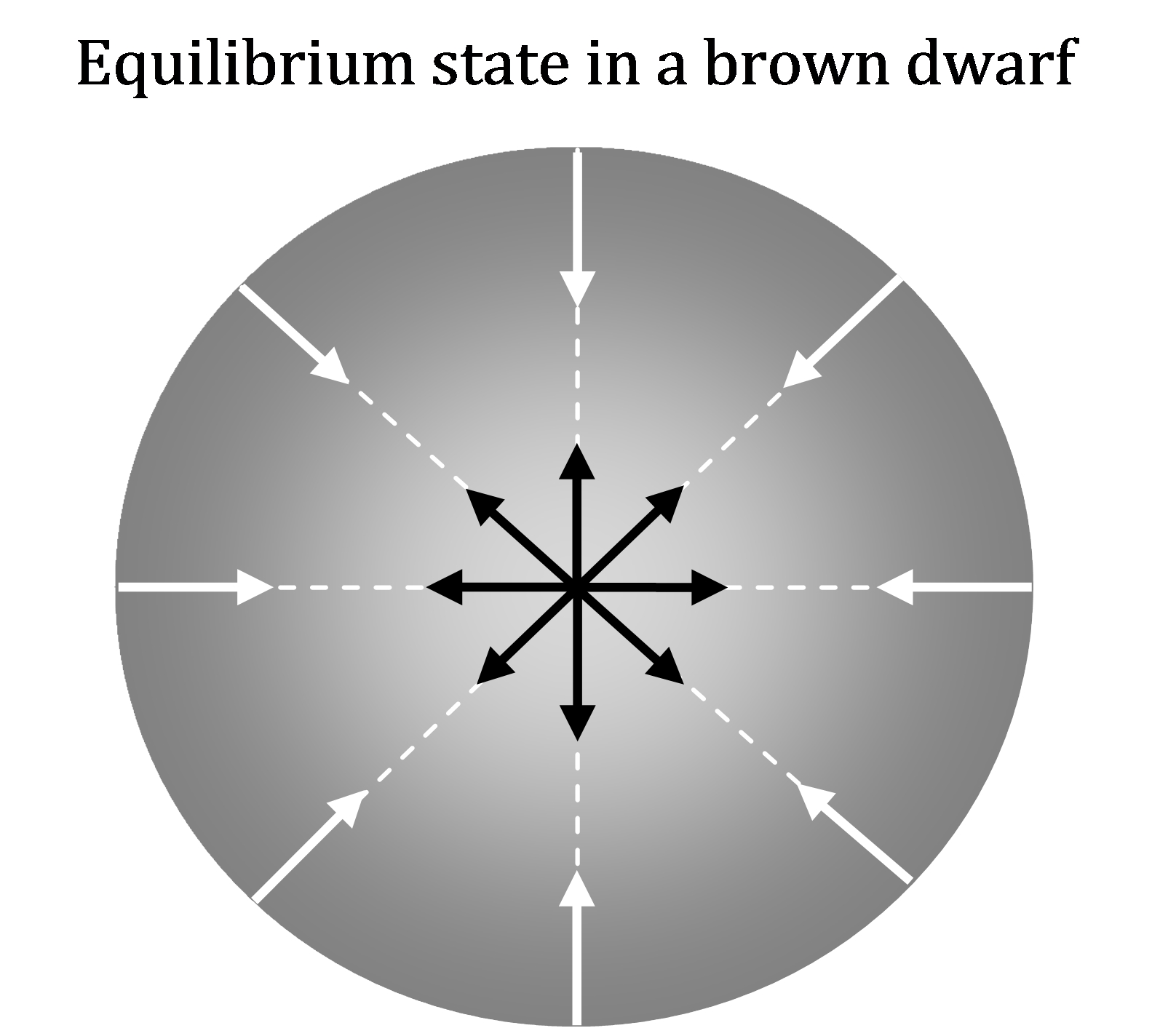}
  \caption{\small A brown dwarf is in equilibrium when the gravitational force (white arrows) is exactly countered by the repulsion between electrons (black arrows) due to Pauli exclusion principle.}
\end{figure}

As can be inferred from above, the upper limit for the mass of a brown dwarf overlaps with the lower limit for the mass of a star. With this idea in mind, our strategy to calculate this limit is to first find an heuristic expression for the mass of equilibrium of a brown dwarf that contains its temperature, and then calculate approximately the value for such mass that corresponds to a temperature of $\sim 10^{7} K$, at which a star is born. This requires an understanding of the role that degeneracy pressure plays in the equilibrium of the brown dwarf. To this effect, some notions developed by the author in another article will be introduced [10].

\section{Exclusion principle and degeneracy pressure} 

The spin is an intrinsic form of angular momentum carried by elementary particles, and is measured in units of the reduced Planck constant $\hbar = h/2 \pi$. Particles with integer spin values ($0\hbar, 1\hbar, 2\hbar,$ etc.) are called \textit{bosons}, whereas particles with fractional values of spin ($\hbar/2, 3\hbar/2, 5\hbar/2,$ etc.) are called \textit{fermions} [6, 7]. The exclusion principle applies only to fermions, which includes the proton, the neutron and the electron. In the case of this class of particles, the exclusion principle establishes that a given region of space can contain a maximum of two electrons with the same energy at the same time [10]. Hence, the exclusion principle acts as a repulsive force between identical fermions when the volume where they are confined is reduced. This repulsive force can be interpreted as an expansive pressure called degeneracy pressure, which opposes to the gravity force. Degeneracy pressure is a phenomenon completely different from coulomb repulsion, because its origin is purely quantum [2, 10].\\

Degeneracy pressure is a direct consequence of \textit{Heisenberg uncertainty principle}, according to which if $\Delta x$ is the uncertainty in the position of the particle, and if $\Delta p$ is the uncertainty in its linear momentum, the minimum value that the product of these quantities can take is:

\begin{equation}
\Delta p \Delta x \approx \hbar
\end{equation}

As the author has shown in detail in another work published in this same journal [10], according to Eq. (1), the minimum volume $V_{min}$ that an object can have is:

\begin{equation} 
V_{min} \approx \frac{1}{2} N_{f} (\Delta x)^{3} \sim N_{f} (\Delta x)^{3}.
\end{equation}

where $(\Delta x)^{3}$ is the smallest volume compatible with the uncertainty principle, and $N_{f}$ is the number of identical fermions contained in $V_{min}$. The physical state of fermions confined in $V_{min}$ is called \textit{degenerate matter}, and is the state adopted by electrons and nuclei in a brown dwarf which has reached equilibrium [10].

\section{The mass of a brown dwarf in equilibrium}

As mentioned above, the first step to calculate $M_{min}$ is to determine the mass of a brown dwarf in equilibrium. In the next lines we develop a simple argument that allows us to obtain an expression of this mass. This argument is based on a simplified semi-classic model of the interior of a brown dwarf. Notwithstanding the simplifications made, the model achieves excellent results, as we will discover in Section 5. As in the heuristic arguments the dimensionless constants have little importance, henceforth we will omit them from the calculations.\\

As a first approach, let us assume that the brown dwarf temperature is low. This assumption allows us to ignore the thermal expansion effects. Thus, we can also assume that gravitational contraction is counteracted by degeneracy pressure alone. Under this conditions, let us consider a brown dwarf of mass $M$ and volume $V_{min}$ composed by $N_{e}$ electrons. According to Eq. (2) we have that:

\begin{equation} 
(\Delta x)^{3} \sim \frac{V_{min}}{N_{e}}.
\end{equation}

Let us consider the cell of minimum size $(\Delta x)^{3}$ inside the brown dwarf, and imagine an electron moving freely inside it, undergoing elastic collisions. If $\Delta p$ is the uncertainty in the electron’s linear momentum and $\Delta t$ is the uncertainty in time, the net force $F_{e}$ applied by the electron over the walls of the cell is of the order of $\Delta p/\Delta t$. Therefore:

\begin{equation} 
F_{e} \sim \frac{\Delta p}{\Delta t} \sim \frac{m_{e} v_{e}}{\Delta t},
\end{equation}

where $m_{e}$ is the electron’s mass and $v_{e}$ it’s velocity. Given that $\Delta t \approx \Delta x/v_{e}$ we get:

\begin{equation} 
F_{e} \sim \frac{m_{e} v_{e}^{2}}{\Delta x}.
\end{equation}

Thus, the pressure $P_{e}$ exerted by the electron is:

\begin{equation} 
P_{e} \sim \frac{F_{e}}{(\Delta x)^{2}} \sim \frac{m_{e} v_{e}^{2}}{(\Delta x)^{3}}.
\end{equation}

Replacing in this equation the value of $(\Delta x)^{3}$ given by Eq. (3):

\begin{equation} 
P_{e} \sim m_{e} v_{e}^{2} \frac{N_{e}}{V_{min}}.
\end{equation}

From Eqs. (1) and (3), $v_{e}$ can be written as:

\begin{equation} 
v_{e} \sim \frac{\hbar}{m_{e}} \frac{1}{\Delta x} = \frac{\hbar}{m_{e}} \left( \frac{N_{e}}{V_{min}} \right)^{1/3}.  
\end{equation}

Introducing in Eq. (7) this value of $v_{e}$ we get:

\begin{equation} 
P_{e} \sim \frac{\hbar^{2}}{m_{e}} \left( \frac{N_{e}}{V_{min}} \right)^{5/3}. 
\end{equation}

Let us define the amount of electrons per unit mass as: 

\begin{equation} 
N=\frac{N_{e}}{M}.
\end{equation}

Based in this definition, Eq. (9) can be written as:

\begin{equation} 
P_{e} \sim \frac{\hbar^{2}}{m_{e}} \left( \frac{NM}{V_{min}} \right)^{5/3} = \frac{\hbar^{2}}{m_{e}} N^{5/3} \left( \frac{M}{V_{min}} \right)^{5/3},
\end{equation}

where $M/V_{min}$ is the mean density. Let us assume that the brown dwarf is composed of a completely ionized hydrogen gas that contains $N_{p}$ protons of mass $m_{p}$ and $N_{e}$ electrons of mass $m_{e}$. Not taking into account the relatively tiny mass of the electrons, we have that $N_{p} \approx M/m_{p}$ . If we also assume that the hydrogen gas has no net electrical charge, it must be that: 

\begin{equation} 
N_{e} = N_{p} \approx \frac{M}{m_{p}}.
\end{equation}

Combining Eqs. (10) and (12):

\begin{equation} 
N \approx \frac{1}{m_{p}}.
\end{equation}

Introducing this value of $N$ into Eq. (11): 

\begin{equation} 
P_{e} \sim \frac{\hbar^{2}}{m_{e} m_{p}^{5/3}} \left( \frac{M}{V_{min}} \right)^{5/3}.
\end{equation}

This expression is the degeneracy pressure $P_{e}$ inside a brown dwarf. From this equation it can be observed that $P_{e}$ does not depend upon temperature, but it’s determined by density alone. If we assume the brown dwarf to be a sphere of radius $R$, and given that $V_{min} \sim R^{3}$, Eq. (14) becomes:

\begin{equation} 
P_{e} \sim \frac{\hbar^{2}}{m_{e} m_{p}^{5/3}} \frac{M^{5/3}}{R^{5}}.
\end{equation}

On the other hand, the gravitational force of the brown dwarf on itself is given by $\sim GM^{2} /R^{2}$, and the surface area has a value $\sim R^{2}$, gravitational pressure will be in the order of: 

\begin{equation} 
P_{g} \sim \frac{F_{g}}{R^{2}} \sim \frac{GM^{2}}{R^{4}}.
\end{equation}

The condition for equilibrium in a brown dwarf is that $P_{e} \approx P_{g}$:

\begin{equation} 
\frac{GM^{2}}{R^{4}} \sim \frac{\hbar^{2}}{m_{e} m_{p}^{5/3}} \frac{M^{5/3}}{R^{5}}. 
\end{equation}

Hence:	

\begin{equation} 
M \sim \frac{\hbar^{6}}{G^{3} m_{e}^{3} m_{p}^{5}} R^{-3}.
\end{equation}

This is the mass for which degeneracy pressure halts the gravitational collapse. However, Eq. (18) does not give us $M_{min}$, because for each value of $R$ there exist a different value for $M$. Under which conditions can be assumed that $M \approx M_{min}$? The key is to add the temperature into the calculations. In effect, if we do not neglect thermal pressure, we conclude that $R$ depends on temperature $T$ and therefore also does $M$. Now, we have that $M \approx M_{min}$ when $T \approx T_{ig}$, where $T_{ig} \approx 10^{7} K$ is the temperature of ignition of thermonuclear reactions. Therefore, to calculate $M_{min}$ we need to find an additional equation that relates $R$ with $T_{ig}$. This is the goal of the next section.

\section{Thermal pressure and the minimum mass required to form a star}

As mentioned above, as the body of gas and dust contracts under its own gravity, a part of the potential energy $E_{P}$ transform into kinetic energy $E_{K}$ [6]. Thus, both forms of energy must be of the same order of magnitude\footnote{It can be shown that, on average, half of the gravitational potential energy transform into kinetic energy and the other half is radiated away according to the \textit{virial theorem}.}: $E_{P} \approx E_{K}$. We can arrive at the same conclusion by noting that if $E_{K}$ were much bigger than $E_{P}$, the particles could easily escape the gravitational pull and a protostar would not form. On the other hand, if $E_{P}$ were much bigger than $E_{K}$, the protostar would collapse under its own weight and could not form either. Therefore we again conclude that $E_{P} \approx E_{K}$.\\

Let us consider a protostar of mass $M$ and radius $R$. The gravitational potential energy (self-energy) of this object is $E_{P} \sim GM^{2} /R$. Thus, according with the conclusion of the preceding paragraph:	

\begin{equation} 
E_{K} \sim \frac{GM^{2}}{R}.
\end{equation}

If the protostar is composed by $N_{p}$ protons of mass $m_{p}$, according to Eq. (12) it follows that $M \approx N_{p} m_{p}$ and thus:

\begin{equation} 
E_{K} \sim \frac{GM(N_{p}m_{p})}{R}.
\end{equation}

If we assume that the protostar is made of an ideal hydrogen gas that is completely ionized, then according to statistical mechanics $E_{K} \sim NkT$, where $N$ is the total number of particles (protons and electrons), $k$ is the Boltzmann constant and $T$ is the absolute temperature. Therefore, Eq. (20) can be written as:   

\begin{equation} 
NkT \sim \frac{GM(N_{p}m_{p})}{R}.
\end{equation}

As we have assumed an electrically neutral hydrogen gas, then the number of electrons $N_{e}$ is equal to the number of protons $N_{p}$. Thus, $N_{e} = N_{p} \approx N$ and Eq. (21) is reduced to:

\begin{equation} 
kT \sim \frac{GMm_{p}}{R}.
\end{equation}

If we assume that $T \approx T_{ig}$, then $M \approx M_{min}$:

\begin{equation} 
R \sim \frac{GM_{min}m_{p}}{kT_{ig}}.
\end{equation}

Replacing this expression of $R$ in Eq. (18):

\begin{equation} 
M_{min} \sim \frac{\hbar^{6}}{G^{3}m_{e}^{3}m_{p}^{5}} \left( \frac{GM_{min}m_{p}}{kT_{ig}} \right)^{-3}. 
\end{equation}

After some algebraic manipulation we finally get the result we set to look for:

\begin{equation} 
M_{min} \sim \left( \frac{\hbar^{2}k}{G^{2}m_{e}m_{p}^{8/3}} \right)^{3/4} T_{ig}^{3/4} \cong 3.08 \times 10^{28} kg \cong 0.016M_{\odot}, 
\end{equation}

where $\hbar = 1.05 \times 10^{-34} J\cdot s, k= 1.38 \times 10^{-23} J\cdot K^{-1}, G= 6.67 \times 10^{-11} N\cdot m^{2}\cdot kg^{-2}, m_{e} = 9.1 \times 10^{-31}kg, m_{p} = 1.67 \times 10^{-27}kg$, and $T_{ig} = 10^{7} K$. We see that $M_{min} \sim 0.016M_{\odot}$ is the same order of magnitude that accepted value, $M_{min} \approx 0.08M_{\odot}$, and therefore it is an excellent approximation. This last value is a spectacular confirmation of the power of theory and observation at work. Indeed, after decades of careful astronomical observations, no star has been detected with a mass lower than $\sim 0.08M_{\odot}$, and therefore neither has a brown dwarf has been detected which mass is above this value [2-4].

\section{Final comments: The Discovery of brown dwarfs and the mínimum mass of stars}
Splitted by a narow gap at $M_{min} \approx 0.08M_{\odot}$, stars and brown dwarfs have a common origin, although their fates are rather different. The former are destined to shine brightly thanks to the ignition of nuclear reactions; the later are destined to slowly cool down. Brown dwarfs are failed stars, therefore, if stars exist brown dwarfs must exist too. However, until mid 1990’s, the validity of star formation models was in serious doubt. The problem was that although models predicted the existence of brown dwarfs, no object of this characteristics was detected by observations. The reason is simple: because of their low temperature, brown dwarfs are very dim, and thus they are difficult to detect [9].\\

We know that the mass of a brown dwarf is less than $\sim 0.08M\odot$, and that its effective temperature is $\sim 10^{3}K$. Is there another distinctive feature that allows to identify a brown dwarf? In the 1990’s a team of astronomers lead by Rafael Rebolo developed the now called lithium test [11], that allows to distinguish a low mass star ($\sim 0.1M_{\odot}$) from a brown dwarf. The gas and dust cloud that forms a protostars contains mainly hydrogen, helium and small amounts of lithium. Low mass stars are entirely convective, their interior is well mixed and the hydrogen, helium and lithium go through the core repeatedly, where they undergo nuclear fusion. Lithium is not as abundant as hydrogen and helium, thus it disappears in a very short time by comparison\footnote{The disappearance of Lithium is produced by the reaction $^{7}L_{i} + ^{1}H \rightarrow 2 ^{4}H_{e}$, which is part of the proton-proton chain.}. However, brown dwarfs are unable to trigger the fusion of lithium, so this element remains intact. Lithium in the atmosphere can be detected observing its particular emission spectra, allowing us to identify brown dwarfs.\\

In 1995 occurred the awaited discovery of a brown dwarf, when a team -also lead by Rafael Rebolo- identified an object which mass was $\sim 0.052M_{\odot}$ and which effective temperature was $\sim2600 K$. But the most important aspect of this discovery is that this object also passed the lithium test. The brown dwarf was named Teide 1, and is located at $\sim 400$ light-years from Earth [12]. This discovery signaled the start of many more brown dwarf detection, every one of which with a mass no more than $\sim 0.08M_{\odot}$. This fact has strengthen the confidence in the stellar formation models and has provided solid evidence in favor of the value $M_{min} \approx 0.08M_{\odot}$. At present time, it is thought that brown dwarfs are almost as abundant as stars.

\section*{Acknowledgments}
I would like to thank to Daniela Balieiro for their valuable comments in the writing of this paper. 

\section*{References}

[1] S.B. Dieterich, et al., Dynamical Masses of $\varepsilon$ Indi B and C: Two Massive Brown Dwarfs at the Edge of the Stellar – substellar Boundary, The Astrophysical Journal, 865 (2018).

\vspace{2mm}

[2] D. Maoz, Astrophysics in a Nutshell, Princeton University Press, Princeton, 2007.

\vspace{2mm}

[3] K.L. Lang, Essential Astrophysics, Springer, Berlin, 2013.

\vspace{2mm}

[4] F. Leblanc, An Introduction to Stellar Astrophysics, Wiley, Chichester, 2010.

\vspace{2mm}

[5] P. Jain, An Introduction to Astronomy and Astrophysics, CRC Press, Boca Raton, 2015.

\vspace{2mm}

[6] K. Krane, Modern Physics, 3 ed., John Wiley and Sons, Hoboken, 2012.

\vspace{2mm}

[7] P.A. Tipler, R.A. Llewellyn, Modern Physics, 6 ed., W. H. Freeman and Company, New York, 2012.

\vspace{2mm}

[8] L. Oster, Astronomía Moderna, Reverté, Barcelona, 2004.

\vspace{2mm}

[9] D. Nardo, The Discovery of Brown Dwarfs, Scientific American, (2000) 27-33.

\vspace{2mm}

[10] J. Pinochet, M. Van Sint Jan, Chandrasekhar Limit: An elementary approach based on Classical Physics and Quantum Theory, Physics Education, 51 035007 (2016).

\vspace{2mm}

[11] R. Rebolo, E.L. Martín, A. Magazzu, Spectroscopy of a brown dwarf in the $\alpha$ persei open cluster, The Astrophysical Journal, 389 (1992) L83-L86.

\vspace{2mm}

[12] R. Rebolo, M.R. Zapatero-Osorio, E.L. Martín, Discovery of a brown dwarf in the Pleiades star cluster, Nature, 377 (1995) 129-131.

\end{document}